\documentclass[11pt,a4paper]{article}
\usepackage{jheppub}
\pdfoutput=1
\usepackage[utf8]{inputenc}
\usepackage{float}
\usepackage{lmodern}
\usepackage{amsmath,amssymb,mhchem}
\usepackage{graphicx}
\usepackage{url}
\usepackage{color}
\usepackage{enumitem}
\usepackage{subfigure}
\usepackage[dvipsnames]{xcolor}
\hypersetup{allcolors=[rgb]{0.0 0.0 0.6},linkcolor=[rgb]{0.75 0.05 0.05}}
\usepackage{bm}
\usepackage{epsfig}
\usepackage{amsmath}
\usepackage{amssymb}
\usepackage{slashed}
\usepackage{color}
\usepackage{accents}
\usepackage{url}
\usepackage[dvipsnames]{xcolor}
\usepackage{cancel}
\usepackage[normalem]{ulem}
\usepackage[colorinlistoftodos]{todonotes}
\usepackage{multirow}

\newcommand{\exclude}[1]{}

\usepackage[normalem]{ulem}

\title{Searching for light neutralinos with a displaced vertex at the LHC}

\author[a,b]{Giovanna Cottin,}
\emailAdd{giovanna.cottin@uai.cl}
\affiliation[a]{Departamento  de  Ciencias,  Facultad  de  Artes  Liberales, \\ Universidad  Adolfo  Ib\'a\~nez,  Diagonal  Las  Torres  2640,  Santiago,  Chile}
\affiliation[b]{Millennium Institute for Subatomic Physics at the High Energy Frontier (SAPHIR), Fern\'andez Concha 700, Santiago, Chile}

\author[c,b]{Juan Carlos Helo,}
\emailAdd{jchelo@userena.cl}
\affiliation[c]{Departamento de F\'{i}sica,
Facultad de Ciencias, Universidad de La Serena,
Avenida Cisternas 1200, La Serena, Chile}

\author[c,b]{Fabi\'an Hern\'andez-Pinto,}
\emailAdd{fhernandezp@alumnosuls.cl}

\author[d]{Nicol\'as A. Neill,}
\emailAdd{naneill@outlook.com}
\affiliation[d]{Departamento de Ingenier\'ia El\'ectrica-Electr\'onica, Universidad de Tarapac\'a, Arica 1010069, Chile}

\author[e,f]{Zeren Simon Wang} 
\emailAdd{wzs@mx.nthu.edu.tw}
\affiliation[e]{Department of Physics, National Tsing Hua University, Hsinchu 300, Taiwan}
\affiliation[f]{Center for Theory and Computation, National Tsing Hua University, Hsinchu 300, Taiwan}

\date{\today}

\abstract{
We study a bino-like light neutralino ($\tilde \chi_1^0$) produced at the LHC
from the decay of a scalar lepton ($\tilde e_L$) through the process $pp\to \tilde e_{L} \to e\tilde \chi_1^0$ in the context of R-parity-violating (RPV) supersymmetry where $\tilde \chi_1^0$ is the lightest supersymmetric particle.
For small masses and RPV couplings, the neutralino is naturally long-lived and its decay products can be identified as displaced tracks.
Following existing searches, we propose a displaced-vertex search strategy for such a light neutralino with a single RPV coupling switched on, $\lambda'_{111}$, in the mass range $10\,\mbox{GeV} \lesssim m_{\tilde \chi_1^0}\lesssim 230\,\mbox{GeV}$.
We perform Monte Carlo simulations
and conclude that at the high-luminosity LHC, the proposed search can probe values of $\lambda'_{111}$ down to two orders of magnitude smaller than current bounds and up to 40 times smaller than projected limits from monolepton searches.
}

\begin{document}

\maketitle
{
  \hypersetup{linkcolor=black}
}

\section{Introduction}

In recent years,
searches for heavy new particles at the Large Hadron Collider (LHC), inspired by supersymmetry (SUSY) and other new physics (NP) scenarios, have yielded no concrete fruit.
On the other hand, the lifetime frontier has become increasingly important.
In particular, the LHC and other high-energy experiments are now actively looking for long-lived particles (LLPs)~\cite{Alimena:2019zri,Lee:2018pag,Curtin:2018mvb}.
After such exotic states are produced, they can travel  macroscopic distances before decaying into Standard Model (SM) or other NP particles, leading to distinctive signatures such as displaced leptons or displaced vertices.
A series of far-detector experiments with the main purpose of searching for LLPs, such as FASER~\cite{Feng:2017uoz} and MoEDAL-MAPP~\cite{Staelens:2019gzt}, have been proposed or approved at different interaction points (IPs) of the LHC.

In various SUSY models, different new particles can be long-lived.
For instance, one can have a long-lived gluino in split SUSY~\cite{Hewett:2004nw}, or a long-lived chargino in compressed SUSY~\cite{Giudice:1998xp,Randall:1998uk}.
In this work, we consider the production of the lightest neutralino in R-parity-violating (RPV) supersymmetry (see Refs.~\cite{Dreiner:1997uz,Barbier:2004ez,Mohapatra:2015fua} for reviews on the model).
The minimal supersymmetric standard model (MSSM) with broken R-parity allows for the lightest supersymmetric particle (LSP) to decay into standard model particles via either bilinear or trilinear RPV couplings.
In this work, we assume that the lightest neutralino is the LSP.

Current bounds on the lightest neutralino mass are much looser than those on the squark, slepton, and gluino masses.
If the GUT-inspired 
relation between the gaugino masses ($M_1\approx 0.5 M_2$) is dropped~\cite{Choudhury:1995pj,Choudhury:1999tn} and dark matter (DM) does not consist of the lightest neutralino~\cite{Belanger:2002nr,Hooper:2002nq,Bottino:2002ry,Belanger:2003wb,AlbornozVasquez:2010nkq,Calibbi:2013poa}, then the lightest neutralino can have $\mathcal O(\mbox{GeV})$ masses or even be massless~\cite{Gogoladze:2002xp,Dreiner:2009ic}.
Such a light neutralino has to be bino-like~\cite{Gogoladze:2002xp,Dreiner:2009ic} and it is in agreement with both astrophysical and cosmological bounds~\cite{Grifols:1988fw,Ellis:1988aa,Lau:1993vf,Dreiner:2003wh,Dreiner:2013tja,Profumo:2008yg,Dreiner:2011fp} as long as it decays (e.g.~in the framework of RPV-SUSY) in order to avoid overclosing the Universe~\cite{Bechtle:2015nua}.
See Ref.~\cite{Domingo:2022emr} for a recent study on the low-energy phenomenomology of a bino-like neutralino lighter than the tau lepton.

Besides the bounds on the lightest neutralino mass, the various RPV couplings are also constrained from both collider and low-energy observables \cite{Allanach:1999ic}.
In this work, we focus on the single coupling $\lambda'_{111}$,
for which the most stringent current bounds stem from neutrinoless double beta decay~\cite{Mohapatra:1986su,Hirsch:1995zi,Hirsch:1995ek,Bolton:2021hje} as well as monolepton searches at the LHC~\cite{ATLAS:2017jbq}.
Details of the neutrinoless double beta decay bounds are described in Sec.~\ref{sec:results}, while a reinterpretation of the monolepton search for our scenario is presented in Sec.~\ref{sec:appendix_reinterpretation}.
Both bounds are shown in Sec.~\ref{sec:results} together with the expected bounds from the displaced vertex (DV) search proposed in this work.
If the coupling $\lambda'_{211}$ is instead switched on, the sensitivity results will be similar but the neutrinoless double beta decay constraints would no longer apply.

If either the lightest neutralino is light or the RPV couplings are small enough, the lightest neutralino becomes long-lived and may be searched for at collider, beam-dump experiments and even atmospheric neutrino detectors~\cite{Candia:2021bsl}.
Phenomenological studies on the exclusion limits in the same scenario already exist from the planned SHiP experiment~\cite{Gorbunov:2015mba,deVries:2015mfw}, proposed LHC far detectors~\cite{Helo:2018qej,Dercks:2018eua,Dercks:2018wum,Dreiner:2020qbi,Dreiner:2022swd,Gehrlein:2021hsk}, Belle II~\cite{Dey:2020juy,Dib:2022ppx}, and future lepton colliders~\cite{Wang:2019orr,Wang:2019xvx}.
These works mainly focus on the lightest neutralino produced from either $B$- or $D$-mesons decays, or $Z$-boson decays, constraining masses below $\sim 45~\text{GeV}$.
In contrast, in this work we consider the on-shell production of a heavy slepton, which then further decays to a charged lepton and a long-lived neutralino.
This allows us to probe neutralino masses up to above 200 GeV, as will be shown in Sec.~\ref{sec:results}.

This work is organized as follows.
We introduce the RPV-MSSM model basics and benchmark scenarios in Sec.~\ref{sec:model}.
The simulation procedure is explained in Sec.~\ref{sec:simulation} together with the description of the event selections.
Numerical results are presented in Sec.~\ref{sec:results} and
we conclude with a summary and conclusions in Sec.~\ref{sec:conclusions}.
Additionally, in Appendix~\ref{sec:appendix_reinterpretation}, we describe our reinterpretation of the LHC monolepton search in the context of the RPV-MSSM.

\section{Model and benchmark scenarios}\label{sec:model}

\begin{figure}[tb]
    \centering
    \includegraphics[width=0.5\textwidth]{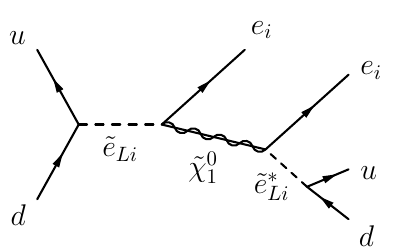}
    \caption{Parton-level Feynman diagram for the neutralino ($\tilde \chi_1^0$) production and decay through the $\lambda'_{i11}$ coupling at the LHC.
    The slepton ($\tilde e_{Li}$) in the left-hand side of the diagram decays into a neutralino and a prompt, charged lepton. 
    We trigger on this prompt lepton while the decay products of the long-lived neutralino are identified as displaced tracks.
    In this scenario, the neutralino is long-lived as a result of its small mass, the off-shell propagator of the heavy slepton ($\tilde e_{Li}^*$), and the small RPV coupling.
    We note that diagrams with a positively charged $s$-channel slepton decaying into the neutralino, or with a neutralino decay into $e^+_i \bar{u} d$, are implied.}
    \label{fig:feyndiag}
\end{figure}

\noindent
In the RPV-MSSM, the MSSM is supplemented with the following RPV superpotential \cite{Weinberg:1981wj,Hall:1983id}
\begin{equation}
    W_{\text{RPV}}
    = \sum_i \mu_i L_i H_u 
    + \sum_{i,j,k} \left(
      \frac{1}{2}\lambda_{ijk} L_i L_j E^c_k
      + \lambda'_{ijk} L_i Q_j D^c_k
      + \frac{1}{2}\lambda''_{ijk} U^c_i D^c_j D^c_k
    \right),\label{eq:WRPV}
\end{equation}
where $Q_i$, $D_i^c$, $U_i^c$, and $L_i$, $E_i^c$ are chiral superfields and $i,j,k=(1,2,3)$ are generation indices.
The $\mu_i$, $\lambda_{ijk}$, and $\lambda'_{ijk}$ couplings violate lepton number ($L$) while the $\lambda''_{ijk}$ couplings violate baryon number ($B$).
If all the RPV terms in Eq.~\eqref{eq:WRPV} are present and unsuppressed, they would allow for a proton decay rate not compatible with current bounds on the proton lifetime\footnote{See Ref.~\cite{Chamoun:2020aft} for a recent study on constraints on RPV couplings from experimental and lattice results of nucleon decays.}.
Therefore, we will consider the scenario where baryon-number-violating couplings ($\lambda''_{ijk}$) are vanishing or negligible.
This can be justified, e.g., by imposing a baryon triality $B_3$ discrete symmetry~\cite{Ibanez:1991pr,Dreiner:2012ae}.
From the remaining $L$-violating terms in Eq.~\eqref{eq:WRPV}, the second trilinear term 
($LQD^c$) allows for the superpartners to be singly produced at the LHC.
In this work we will focus on the $LQD^c$ term assuming all the other RPV couplings are zero. The Yukawa couplings generated by this operator are
\begin{align}
    L_{\text{RPV}} = \lambda'_{ijk}
    \left(
      \tilde \nu_{i L} \bar d_{k R} d_{j L}
      + \tilde d_{j L} \bar d_{k R} \nu_{i L}
      + \tilde d^*_{k R} \bar \nu^c_{iR} d_{j L}
      - \tilde e_{i L} \bar d_{kR} u_{j L}
    \right.\ \ \ \nonumber\\
    \left.
      - \tilde u_{j L} \bar d_{kR} e_{i L}
      - \tilde d^*_{k R} \bar e^c_{iR} u_{j L}
    \right) + h.c.\label{eq:LRPV}
\end{align}
The fourth term in Eq.~\eqref{eq:LRPV}, which includes a charged slepton, allows for the neutralinos to be produced at the LHC together with a prompt charged lepton, as shown in  Fig.~\ref{fig:feyndiag}.
As it will be explained in Sec.~\ref{sec:simulation}, this prompt lepton will be used as a trigger in the proposed search.
We consider $\tilde \chi_1^0$ to be the lightest supersymmetric particle, so it can only decay through the RPV terms, with a total decay width proportional to the $\lambda_{ijk}^{'}$ couplings squared,
\begin{equation}
    \Gamma_{\tilde \chi^0_1} \propto m_{\tilde \chi^0_1}^5\left(\frac{\lambda_{ijk}^{'}}{m_{\tilde f}^2}\right)^2,
    \label{Gamma}
\end{equation}
where $m_{\tilde f}$ is the mass of the corresponding sfermion mediating the decay.
Consequently, for small enough RPV couplings (and $m_{\tilde \chi_1^0}$) the neutralino will be long-lived.
The decay products of this long-lived neutralino (shown in Fig.~\ref{fig:feyndiag}) will be identified as displaced tracks from a common origin, i.e., a displaced vertex (DV).

From the 27 flavor combinations of the $\lambda'_{ijk}$ couplings, the strongest sensitivity at the LHC will be for $\lambda'_{i11}$ ($i=1,2$) as a result of the proton parton distribution functions and higher reconstruction efficiencies for electrons and muons compared to tau leptons.
For concreteness, in this work we analyze the sensitivity of the LHC to $\lambda'_{111}$, but the search strategy described in this work is expected to give similar constraints for $\lambda'_{211}$.
For simplicity, all the superpartners different from $\tilde \chi_1^0$ and $m_{\tilde e_L}$ are taken to be heavy (10 TeV), so that they are effectively decoupled.
Although this mass hierarchy may be difficult to achieve in a realistic model, this is a phenomenological consideration chosen to define our benchmarks, so the phenomenology at the LHC can be controlled by the following three parameters only: 
\begin{equation}
\lambda'_{111}, m_{\tilde e_L}, m_{\tilde \chi^0_1}.
\end{equation}

In Sec.~\ref{sec:results} we will present our numerical results with four benchmark scenarios of the selectron mass, $m_{\tilde e_L}$: $1\mbox{ TeV}$, $2.5\mbox{ TeV}$, $5\mbox{ TeV}$, and $7\mbox{ TeV}$, while varying the mass of the lightest neutralino ($m_{\tilde \chi^0_1}$) and $\lambda'_{111}$ freely.

\section{Simulation and event selection}\label{sec:simulation}

Inspired by the ATLAS 13-TeV SUSY search for displaced vertices~\cite{Aaboud:2017iio}, we focus on a search strategy that identifies the $\tilde{\chi}^{0}_{1}$ decay products inside the inner tracker as displaced tracks, which can come from the hadronized quarks or the displaced electron from the neutralino decay.

\begin{figure}[tb!]
    \centering
    \includegraphics[scale=0.75]{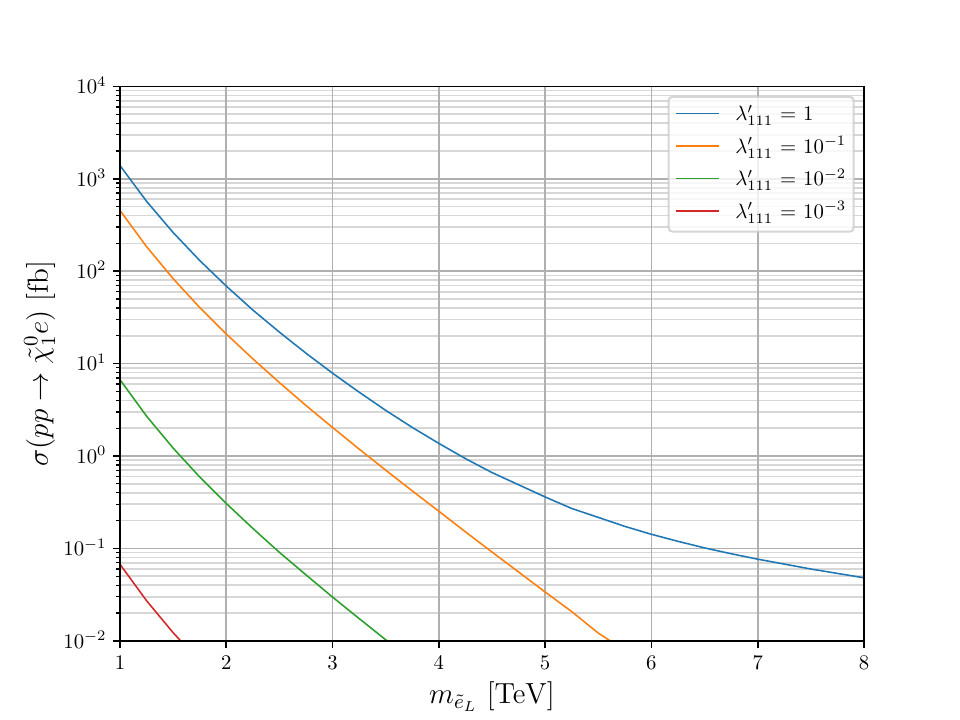}
    \caption{Production cross section of $pp\rightarrow \tilde{\chi}^{0}_{1} e$ for different values of $\lambda'_{111}$ as a function of $m_{\tilde{e}_L}$ for $m_{\tilde{\chi}^0_1}=$100 GeV.}
    \label{fig:xsec}
\end{figure}

We use the RPV-MSSM UFO model file implemented in Ref.~{\cite{RPVUFO}},  with flavor diagonal mixing matrices for sfermions. We also set the lightest neutralino to be a pure bino in the model spectrum.
We simulate the process $pp\rightarrow \tilde{\chi}^{0}_{1} e$ in MadGraph 5~\cite{Alwall:2011uj} at $\sqrt{s}=13$ TeV, and generate parton-level LHE events with displaced information. The decay widths of the selectron and the lightest neutralino are automatically computed by MadGraph 5.
Figure~\ref{fig:xsec} shows the production cross section of $pp\to \tilde{\chi}^0_1 e$ for different values of $\lambda'_{111}$ as a function of the selectron mass, for a fixed neutralino mass of 100 GeV.
For other neutralino mass values below $\sim 250$ GeV, there are no appreciable differences, so we take only one benchmark neutralino mass of 100 GeV here.
We note that for Figure~\ref{fig:xsec} and the remaining simulation,
we set the kinematic cuts of $p_{T} \geq 10$ GeV  and $|\eta|>2.5$ for the outgoing electron or positron at the generation level.

Events are further read within Pythia 8~\cite{Sjostrand:2014zea} for showering and hadronization. We then perform a custom detector simulation within Pythia 8 for electrons, and displaced tracks and vertices. We start by selecting events triggering on a prompt, isolated electron with $p_{T}>25$ GeV and with $|\eta| < 2.47$. Displaced vertices are then selected from tracks with a high transverse impact parameter, of $|d_{0}|> 2$ mm and $p_{T} > 1$ GeV. Vertices are required to be within the inner tracker acceptance, with transverse decay positions $r_{\text{DV}}$ between 4 and 300 mm, as well as longitudinal distance  $|z_{\text{DV}}|< 300$ mm. Additionally, displaced vertices must have at least 5 tracks and have an invariant mass  $m_{\text{DV}} \geq 10$ GeV (for which we assume all the tracks have the mass of the pion). These last two cuts define the region where signal is expected to be found free of Standard Model and instrumental background events~\cite{Aaboud:2017iio}. In order to further characterize the detector response to displaced vertices within the above mentioned regions, we also make use of the parametrized vertex-level efficiencies provided by ATLAS in Ref.~\cite{Aaboud:2017iio}. A similar search for a long-lived right-handed neutrino in the context of a left-right symmetric model was performed in Refs.~\cite{Cottin:2019drg,Cottin:2018kmq}.

\begin{figure}[tb!]
    \centering
    \includegraphics[scale=0.75]{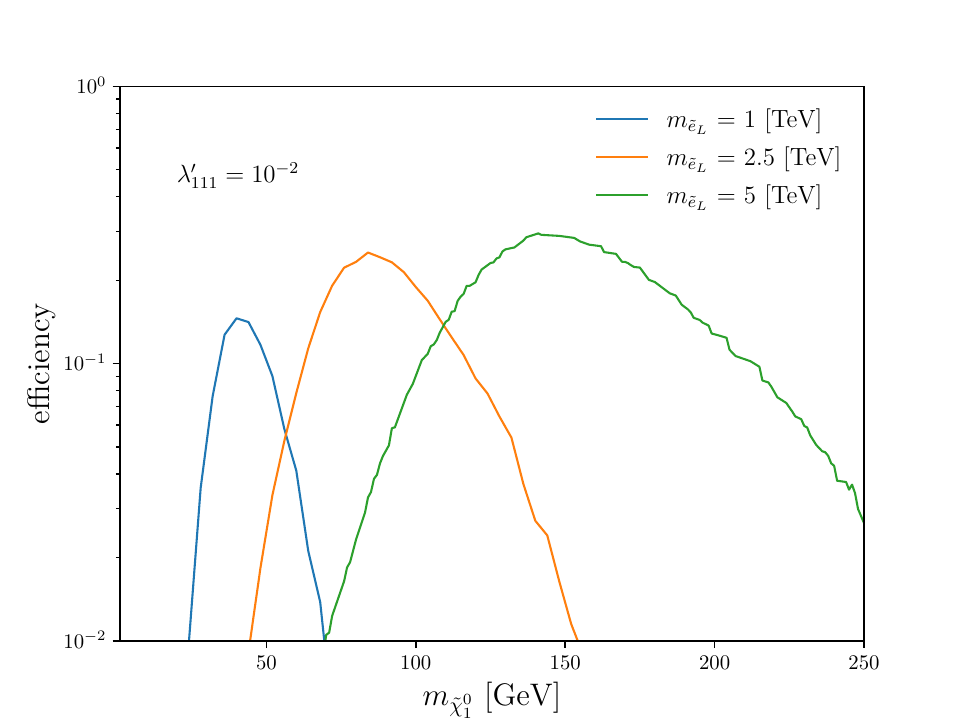}
    \caption{DV selection efficiency as a function of the neutralino mass, for three values of the slepton mass. Here as a sample scenario, we fix $\lambda'_{111}=10^{-2}$.}
    \label{fig:DVvsmass}
\end{figure}

In figure \ref{fig:DVvsmass} we show the overall selection efficiency of our DV strategy as a function of the neutralino mass, for three benchmark values of the slepton mass $m_{\tilde{e}_L}$: 1, 2.5 and 5 TeV, and a fixed coupling, $\lambda'_{111}=10^{-2}$.
We observe that the larger is the slepton mass, the higher is the peak value in efficiency at a higher value of $m_{\tilde{\chi}^0_1}$.
The highest efficiency is achieved at values of the boosted decay length of the neutralino,
\begin{equation}
    \beta\gamma c\tau = \frac{p}{m_{\tilde{\chi}^0_1}}\frac{\hbar c}{\Gamma_{\tilde{\chi}^0_1}},
\end{equation}
that lie within a certain optimal range (of order $\mathcal{O}$(cm)) corresponding to the ATLAS inner detector geometry and predict the largest decay probability inside the ATLAS fiducial volume.
Here, $p$ denotes the 3-momentum magnitude of the light neutralino.
The neutralino boosted decay length is proportional to $m^4_{\tilde{e}_L}/m^6_{\tilde{\chi}_1^0}$ (see Eq.~\eqref{Gamma}).
As a result, with $\lambda'_{111}$ fixed, the peak efficiency is obtained at a larger $m_{\tilde{\chi}^0_1}$ for a heavier slepton mass.
Moreover, for heavier slepton and neutralino masses, the prompt electron tends to be harder, the displaced tracks have a larger transverse impact parameter as well as the transverse momentum, and the displaced vertices consist of a larger number of tracks and a heavier $m_{\text{DV}}$; altogether these lead to a better overall efficiency at the peak.

\begin{figure}[ht!]
    \centering
    \includegraphics{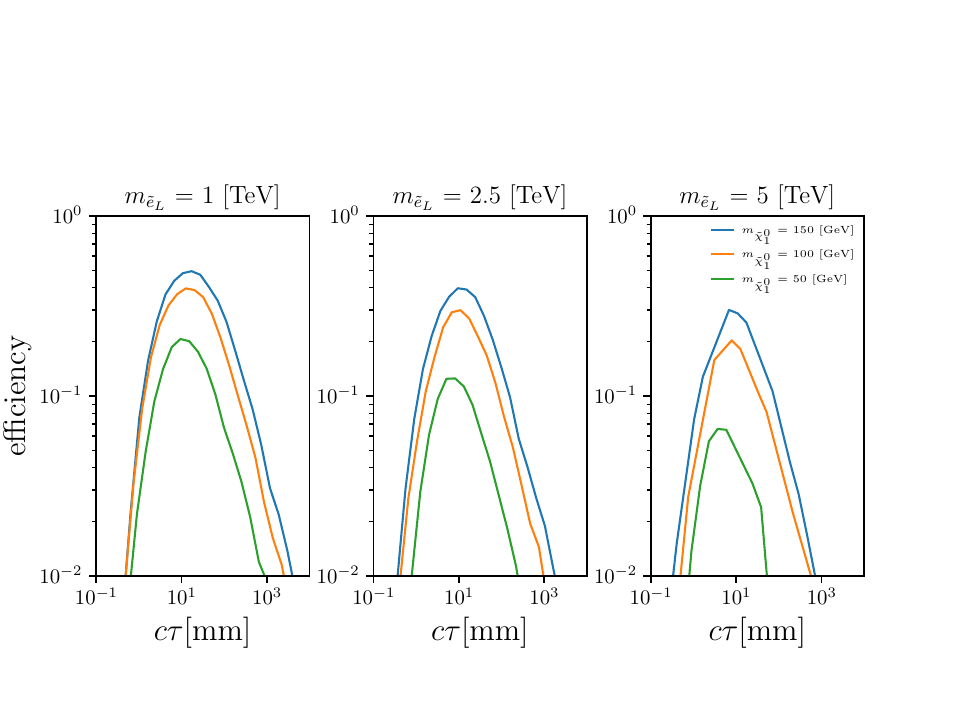}
    \caption{DV selection efficiency as a function of the c$\tau$ of the neutralino, for three fixed slepton masses and three fixed values of the neutralino mass.  }
     \label{fig:DVvsctau}
\end{figure}

This last feature of high mass displaced vertices with higher number of tracks can also be seen in figure \ref{fig:DVvsctau}, which shows the DV efficiency as a function of the neutralino proper decay length, for representative benchmarks with fixed slepton masses, $m_{\tilde{e}_L}$: 1, 2.5 and 5 TeV, and fixed neutralino masses, $m_{\tilde{\chi}^0_1}$: 50, 100 and 150 GeV. Independent of the value of $m_{\tilde{e}_L}$, the larger the neutralino mass, the higher is the peak efficiency in lifetime.

With the above DV search strategy, we can then estimate $95\% $ confidence level (C.L.) exclusion limits under the assumption of zero background, in the RPV coupling and neutralino mass planes in the following section.

\section{Results}\label{sec:results}

\begin{figure}[ht!]
    \centering
    \includegraphics[scale=0.85]{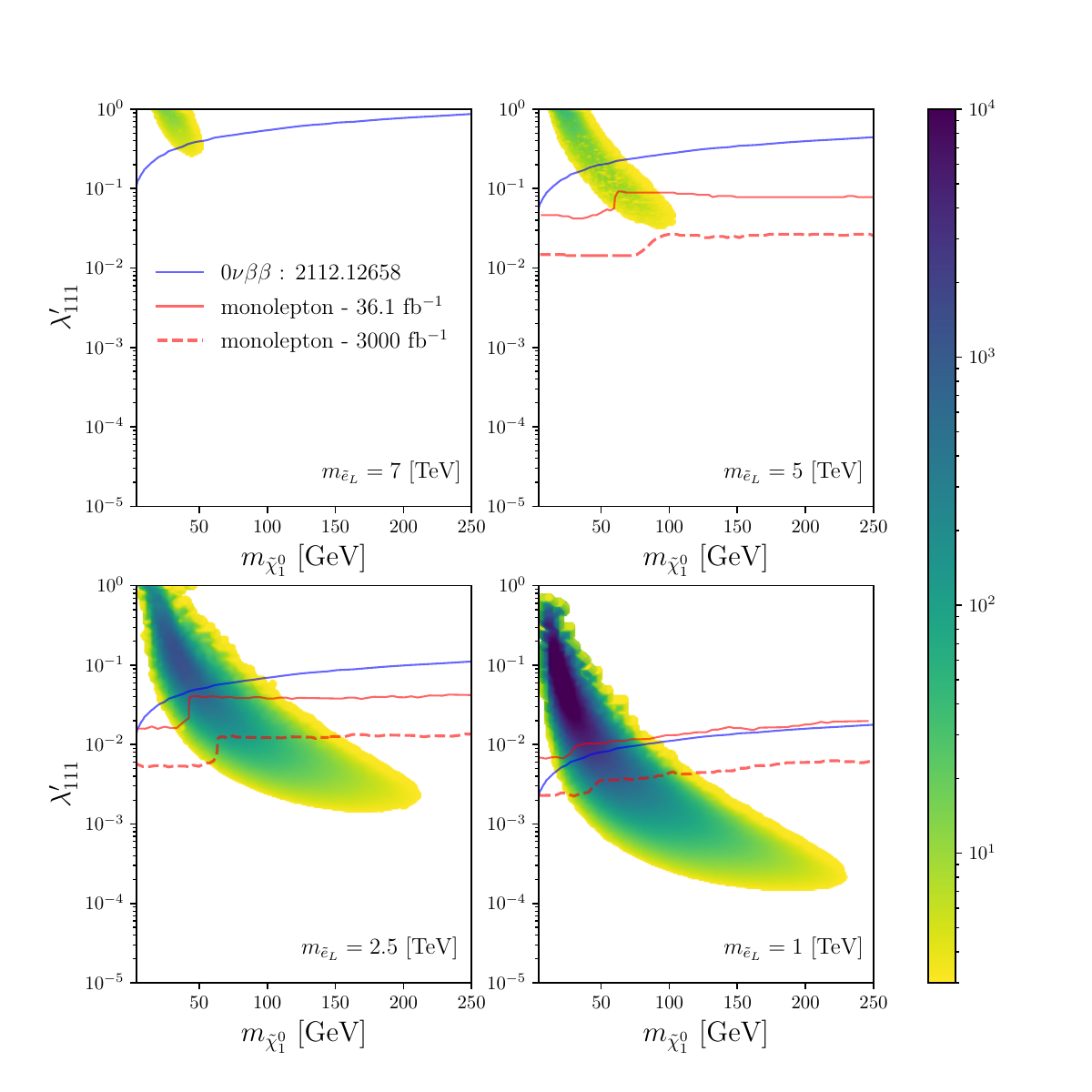}
    \caption{Sensitivity reach in the  $\lambda'_{111}$ vs.~$m_{\tilde \chi_1^0}$ plane with the proposed DV search strategy with 3000 fb$^{-1}$. The number of expected signal events depicted in the color bar starts from 3. Red lines show the recasted monolepton limits for 36.1 fb$^{-1}$ (solid red line) and our high-luminosity LHC projection to 3000 fb$^{-1}$  (dashed red line). Constraints from $0\nu\beta\beta$ are shown in solid blue. 
    }
    \label{fig:limits_final}
\end{figure}

We proceed to calculate the sensitivity to the trilinear RPV coupling $\lambda'_{111}$ with our DV search strategy for long-lived light neutralinos at the ATLAS inner tracker detector. For simplicity, we will assume that the $\lambda'_{111}$ coupling is the only nonzero RPV coupling, and that the masses of the squarks are heavy enough to be outside of LHC range, except for the selectron mass, $m_{\tilde e_L}$.

 In figure \ref{fig:limits_final} we show the expected sensitivity for light neutralinos with our DV search strategy for a luminosity of 3000 fb$^{-1}$.  The $95\% $ C.L. exclusion limits are displayed in the $\lambda'_{111}$ vs.~$m_{\tilde \chi_1^0}$  plane for selectron masses $m_{\tilde e_L} = (1, 2.5, 5, 7)$ TeV. This is so as we are assuming a zero background search and requiring three signal events.

 In all benchmarks, the sensitivity is mostly limited by neutralinos decaying far away beyond the trackers (towards the lower left of the colored contours) and neutralinos decaying too promptly (towards the upper right of the colored contours). The reach in the neutralino mass is lower for higher selectron masses, as it is limited by production cross-section. With this DV search at 3000 fb$^{-1}$, we are able to exclude $\lambda'_{111}$ values as low as $\sim 10^{-4}$ for $m_{\tilde \chi_1^0} \sim 230$ GeV, and $m_{\tilde e_L}=1$ TeV.
 
  Figure \ref{fig:limits_final} also compares our limits with current constraints from neutrinoless double beta decay searches at GERDA \cite{GERDA:2020xhi} and monolepton searches at ATLAS~\cite{ATLAS:2017jbq} with $36.1\,\mbox{fb}^{-1}$ of integrated luminosity.  The neutrinoless double beta decay limits were obtained by comparing the theoretical calculations from Ref.~\cite{Bolton:2021hje} of the RPV SUSY contribution to $0\nu\beta\beta$ half-life, mediated by light neutralinos and selectrons\footnote{Figure 3 in Ref. \cite{Bolton:2021hje} shows the calculation of the contribution of light neutralinos and selectrons to the $0\nu\beta\beta$ half-life for $\lambda'_{111} = 10^{-3}$ and $m_{{\tilde e}_L}$ = 2 TeV (green dashed line). We have re-scaled these results for different values of $\lambda'_{111}$ and $m_{\tilde e_L}$ using the proportionality relation $(T_{1/2}^{0\nu\beta\beta})^{-1}  \propto |\lambda_{111}^{'2}/{m_{\tilde{e}_L}^4}|^2$ (see equation (4.5) in Ref. \cite{Bolton:2021hje}).} (see figure~\ref{fig:feyndiag} and rotate it by 90 degrees clockwise to visualize the Feynman diagram contributing to $0\nu\beta\beta$ decay) with the experimental current limits on the $0\nu\beta\beta$ half-life $T_{1/2}^{exp} > 1.8 \times 10^{26}$yr for the isotope $^{76}$Ge. 
 
 On the other hand, the monolepton limits are based on our reinterpretation of the ATLAS search described in Appendix~\ref{sec:appendix_reinterpretation}, which corresponds to the red solid curve in figure~\ref{fig:limits_final}. We obtain the limits by extracting a contour on the significance at $36.1\,\mbox{fb}^{-1}$, i.e. $Z_{36} \equiv S/\sqrt{B}$, with $S$ the number of signal events after the monolopton cuts and $B$ the number of background events taken from the ATLAS search, at $Z_{36}=2$.    The dashed red curve in figure~\ref{fig:limits_final} is our projected limit for the same monolepton search but for a luminosity of 3000 fb$^{-1}$. The contour is again obtained at $Z_{3000}=2$ after re-scaling with $Z_{3000}\equiv \sqrt{3000/36.1} \cdot Z_{36}$.

For higher selectron mass, the monolepton limits become less stringent, owing to the smaller production cross-sections. We also note a transition in the monolepton search sensitivity happening as the mass of the neutralino increases and ceases to be long-lived, see figure~\ref{fig:limits_final}. This is understood as follows. Monolepton searches are efficient, provided the electron $p_{T}$ and the missing transverse momenta are high enough for the events to pass the cut on transverse mass (see Appendix~\ref{sec:appendix_reinterpretation}).  When the mass of the neutralino is small enough for it to decay outside the detector, all of its momentum contributes to missing transverse momenta (as opposed to only a fraction when it has visible decay products inside the detector). As a result, missing transverse momenta is high enough for the event selections to be efficient when the neutralino is long-lived. For a more prompt neutralino, the monolepton signal efficiencies decrease,  remaining constant for a large part of the neutralino masses thanks to the contributions to the missing transverse momenta coming from prompt activity.

\section{Conclusions}\label{sec:conclusions}
In recent years, increasingly more searches for long-lived particles have been proposed and performed at the LHC and other experiments.
In R-parity-violating supersymmetry (RPV-SUSY), the lightest neutralino is allowed to be light with mass in the GeV scale, as long as it decays via RPV couplings.
Such a light neutralino also must be dominantly bino-like.
If both the mass and the non-vanishing RPV couplings are small, the lightest neutralino is naturally long-lived.

In this work, we have proposed a search strategy based on an existing ATLAS 13-TeV SUSY search, and performed Monte-Carlo simulations to estimate the sensitivities at the high-luminosity LHC (HL-LHC) to such a light neutralino with a single RPV coupling switched on, $\lambda'_{111}$.
We consider on-shell production of an selectron from $pp$ collisions via $\lambda'_{111}$ which then decays promptly into an electron and the lightest neutralino.
The lightest neutralino travels a macroscopic distance before decaying into an electron and two quarks via the same RPV coupling and an off-shell selectron.
Thus, this theoretical scenario comes with only three free parameters: the RPV coupling $\lambda'_{111}$, the neutralino mass $m_{\tilde{\chi}_1^0}$, and the selectron mass $m_{\tilde{e}}$.

For numerical results, we present plots of search efficiencies as functions of either the neutralino mass or its proper decay length, $c\tau$, as well as plots of final sensitivities in the plane $\lambda'_{111}$ vs.~$m_{\tilde{\chi}_1^0}$ for four benchmark selectron masses, $m_{\tilde{e}_L}$: 1, 2.5, 5, and 7 TeV.
Our final results show that for 1 TeV selectron mass, the proposed search at the HL-LHC can probe values of $\lambda'_{111}$ up to two orders of magnitude
smaller than current bounds from neutrinoless double beta decay experiments, as well as up to 40 times smaller than our recast of an LHC monolepton search with an integrated luminosity of 36.1 fb$^{-1}$, projected to the final HL-LHC target, 3000 fb$^{-1}$, for neutralino masses between 10 GeV and 230 GeV.
However, for a heavy selectron of mass 7 TeV, our sensitivities are rather limited, and at most comparable with the bounds from neutrinoless double beta decay at $m_{\tilde{\chi}^0_1}\sim 50$ GeV.

We further note that while we have focused on the single coupling $\lambda'_{111}$, our results are almost equally applicable to the same scenario but with another RPV coupling, $\lambda'_{211}$, as at the LHC, we expect prompt muon efficiencies to be similar from that of electrons. For the case of a single coupling $\lambda'_{211}$ the limits from neutrinoless double beta decay do not apply.

\acknowledgments
We thank Benjamin Fuks and Torbj\"orn Sj\"ostrand for useful discussions on the UFO implementation and Pythia 8, respectively.
G.C. acknowledges support from ANID FONDECYT grant No. 11220237.
G.C., J.C.H. and F.H.P. also acknowledge support from grants ANID FONDECYT No. 1201673 and ANID – Millennium Science Initiative Program ICN2019\_044. J.C.H. acknowledges the financial support of DIDULS/ULS, through the project PTE202135.
Z.S.W. is supported by the Ministry of Science and Technology (MoST) of Taiwan with grant number MoST-110-2811-M-007-542-MY3.
N.A.N. was supported by ANID (Chile) under the grant ANID REC Convocatoria Nacional Subvenci\'on a Instalaci\'on en la Academia Convocatoria A\~no 2020, PAI77200092.

We thank Wei Liu for pointing out an error in Figure~\ref{fig:xsec} in the previous version of the article.

\clearpage

\appendix
\section{Reinterpretation of monolepton search}\label{sec:appendix_reinterpretation}

The ATLAS collaboration has presented a search for a monolepton signal based on $\mathcal{L} = 36.1$ fb$^{-1}$ of statistics taken at $\sqrt{s}=13$ TeV~\cite{ATLAS:2017jbq}. We reinterpret 
this existing prompt search for a new $W'$ gauge boson decaying to an electron and a neutrino, $W'\rightarrow e \nu$, in the context of our RPV 
signal, $\tilde{e}\rightarrow e \tilde{\chi}^{0}_{1}$. 

For our recast, we first validated the $W'$ Sequential Standard Model (SSM) signal model~\cite{Altarelli:1989ff}. The simulation for the $W'$ signal was done with Pythia 8~\cite{Sjostrand:2014zea} for both production and decay. We performed a custom detector simulation where electrons are reconstructed as an isolated prompt object within $|\eta| < 2.47$ with smeared momenta (idem as in Ref.~\cite{Cottin:2018nms}). Missing transverse momenta, $p^{miss}_{T}$, is reconstructed from all visible physics objects (following Ref.~\cite{Allanach:2016pam}, with a standard reconstruction that includes a vector sum of the $p_{T}$ of jets, leptons and unclustered deposits of energy not associated to leptons and jets). Everything decaying outside the inner detector is considered stable. We consider a cylinder with inner detector dimensions radius $r= 1100$ mm and length $|z|= 2800$ mm, as in Ref.~\cite{Allanach:2016pam}. The following cuts are applied:

\begin{itemize}
\item {One electron with $p_{T} > 120$ GeV and transverse energy $E_{T}> 4.5$ GeV}
\item {Missing transverse momenta $p^{miss}_{T} > 65$ GeV}
\item {Transverse mass $m_{T}=\sqrt{2 p_{T} \cdot p^{miss}_{T} (1-\cos{\Delta\phi})} > 130$ GeV. Here $p_{T}$ corresponds to the transverse momentum of the electron, $p^{miss}_{T}$ is the missing transverse momenta, and $\Delta\phi$ is the azimuthal angle between these vectors. }
\end{itemize}

\begin{figure}[ht!]
    \includegraphics[width=0.5\textwidth,angle=0]{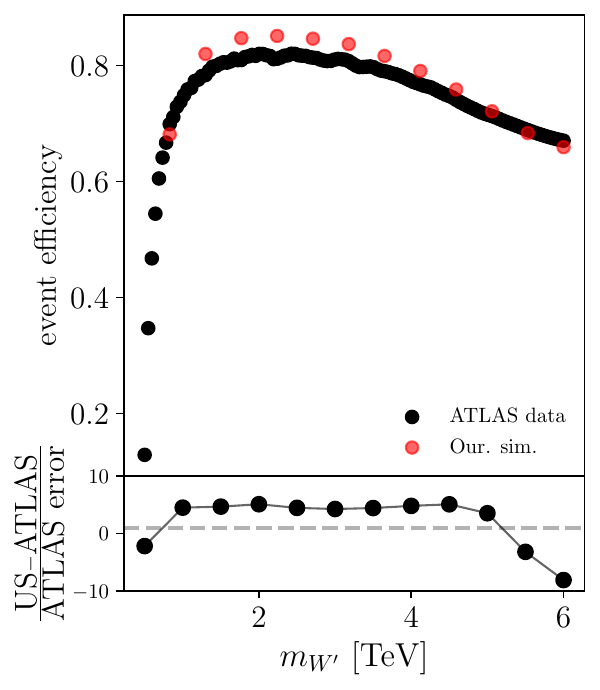}
    \includegraphics[width=0.5\textwidth,angle=0]{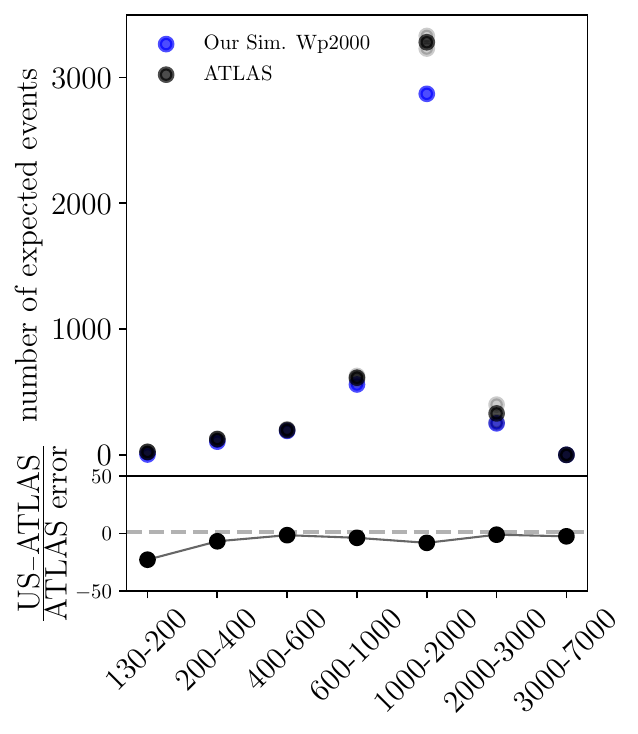}
    \caption{Validation plots per $m_{W'}$ mass point and per $m_{T}$ ATLAS-defined bins. Comparison with the ATLAS expectation is provided in the bottom frames. (left)  Event efficiency as a function of $m_{W'}$. (right) Number of expected events as a function of $m_{T}$ bins.}
    \label{monoleptonValidation}
    \end{figure}

A plot of the overall signal efficiency after all cuts as a function of $W'$ mass is shown in the left frame of figure~\ref{monoleptonValidation}, compared with the ATLAS auxiliary figure 6 from HepData. We also validate Table 1 of the ATLAS paper~\cite{ATLAS:2017jbq}, and extract the number of expected events from $m_{T}$ distributions, which we generate for each $m_{W'}$ mass point. These $m_{T}$ distributions are given after all cuts in the ATLAS paper, so we reproduce these histograms in order to extract event level efficiencies. We calculate the signal efficiency (for a given mass benchmark) after integrating over the $m_{T}$ bins defined by ATLAS. An example benchmark for $m_{W'}=2$ TeV is shown in the right frame of figure~\ref{monoleptonValidation}. In general, most bins match within $30\%$. The biggest discrepancy is found to be for the first $m_{T}$ bin of [130-200] GeV for the $m_{W'}=4$ TeV benchmark, which reaches $55\%$.

We implement the same above mentioned strategy to our RPV signal. As for calculating the monolepton exclusion in figure~\ref{fig:limits_final}, we take the background numbers directly from the ATLAS paper~\cite{ATLAS:2017jbq}, which are provided in HepData in bins of $m_{T}$. We note that monolepton searches have been reinterpreted in a similar way (but without considering detector effects) in Ref.~\cite{Nemevsek:2018bbt} in the context of right-handed $W_{R}$ bosons and heavy neutrinos.

\clearpage

\bibliography{main}
\bibliographystyle{apsrev}
\end{document}